\documentclass[prb,twocolumn,superscriptaddress,secnumarabic,showpacs,amssymb,aps]{revtex4-1}
\usepackage{graphicx,amsmath,bm,color,natbib}

\begin{document}
\title{\mre{Anisotropy and multiband superconductivity in {\SRO}}}
\author{S.~J.~Kuhn}
\altaffiliation{Current address: Department of Physics, Duke University,  Durham, NC 27708}
\affiliation{Department of Physics, University of Notre Dame, Notre Dame, IN 46556}
\author{W.~Morgenlander}
\affiliation{Department of Physics, University of Notre Dame, Notre Dame, IN 46556}
\author{E.~R.~Louden}
\affiliation{Department of Physics, University of Notre Dame, Notre Dame, IN 46556}
\author{C.~Rastovski}
\affiliation{Department of Physics, University of Notre Dame, Notre Dame, IN 46556}
\author{W.~J.~Gannon}
\altaffiliation{Current address: Department of Physics and Astronomy, Texas A \& M University,  College Station, TX 77843}
\affiliation{Department of Physics and Astronomy, Northwestern University,  Evanston, IL 60208}
\author{H.~Takatsu}
\altaffiliation{Current address: Department of Energy and Hydrocarbon Chemistry, Graduate School of Engineering, Kyoto University, Kyoto 615-8510, Japan}
\affiliation{Department of Physics, Graduate School of Science, Kyoto University, Kyoto, 606-8502, Japan}
\author{D.~C.~Peets}
\altaffiliation{Current address: Department of Physics, Fudan University, Shanghai, China 200433}
\affiliation{Department of Physics, Graduate School of Science, Kyoto University, Kyoto, 606-8502, Japan}
\author{Y.~Maeno}
\affiliation{Department of Physics, Graduate School of Science, Kyoto University, Kyoto, 606-8502, Japan}
\author{C.~D.~Dewhurst}
\affiliation{Institut Laue-Langevin, 6 Rue Jules Horowitz, F-38042 Grenoble, France}
\author{J.~Gavilano}
\affiliation{Laboratory for Neutron Scattering, Paul Scherrer Institute, CH-5232 Villigen, Switzerland}
\author{M.~R.~Eskildsen}
\email{Corresponding author: eskildsen@nd.edu}
\affiliation{Department of Physics, University of Notre Dame, Notre Dame, IN 46556}
\date{\today}

%macros
\newcommand{\SRO}{Sr$_2$RuO$_4$}
\newcommand{\KFA}{KFe$_2$As$_2$}
\newcommand{\Tc}{\ensuremath{T_{\text{c}}}}
\newcommand{\kB}{k_B}
\newcommand{\Gac}{\ensuremath{\Gamma_{ac}}}
\newcommand{\GVL}{\ensuremath{\Gamma_{\text{VL}}}}
\newcommand{\GHcii}{\ensuremath{\Gamma_{Hc2}}}
\newcommand{\Hcii}{\ensuremath{H_{\text{c2}}}}
\newcommand{\QVL}{\ensuremath{Q_{\text{VL}}}}
\newcommand{\QVLvec}{\bf{Q_{\text{VL}}}}
\newcommand{\fq}{\ensuremath{\Phi_0}}

\newcommand{\mre}[1] {\textcolor{black}{#1}}
\newcommand{\sjk}[1] {\textcolor{black}{#1}}

\begin{abstract}
Despite numerous studies the exact nature of the order parameter in superconducting {\SRO} remains unresolved.
We have extended previous small-angle neutron scattering studies of the vortex lattice in this material \mre{to a wider field range, higher temperatures,
and with the field applied close to both the $\langle 100 \rangle$ and $\langle 110 \rangle$ basal plane directions.}
\mre{Measurements at high field were made possible by the use of both spin polarization and analysis to improve the signal-to-noise ratio.}
\mre{Rotating the field towards the basal plane causes a distortion of the square vortex lattice observed for  $\bm{H} \parallel \langle 001 \rangle$,
and also a symmetry change to a distorted triangular symmetry for fields close to $\langle 100 \rangle$.}
\mre{The vortex lattice distortion allows}
%This allow
us to determine the intrinsic superconducting anisotropy between the $c$ axis and the Ru-O basal plane, yielding a value of $\sim 60$ at low temperature and low to intermediate fields.
This greatly exceeds the upper critical field anisotropy of $\sim 20$ at low temperature, \mre{reminiscent of} Pauli limiting.
Indirect evidence for Pauli paramagnetic effects on the unpaired quasiparticles in the vortex cores are observed, but a direct detection lies below the measurement sensitivity.
The superconducting anisotropy is found to be independent of temperature but increases for fields $\gtrsim 1$~T, indicating multiband superconductvity in {\SRO}.
Finally, the temperature dependence of the scattered intensity provides further support for gap nodes or deep minima in the \mre{superconducting gap}.
\end{abstract}


\maketitle

%% INTRODUCTION
\section{Introduction}
\label{Intro}
The superconducting state emerges due to the formation and condensation of Cooper pairs,
although the exact microscopic mechanism responsible for the pairing in different materials varies and in many cases remains elusive.
In the %prominent
case of strontium ruthenate,
multiple experimental and theoretical studies provide compelling %support
\mre{evidence} for triplet pairing of carriers (electrons and/or holes) and an odd-parity, $p$-wave order parameter,\cite{Mackenzie:2003wp,Maeno:2012ew}
\mre{which would classify {\SRO} as an intrinsic topological superconductor.\cite{Sato:2017th}}
%Namely,
\mre{This is supported by} $\mu$SR,\cite{Luke:1998wm} Josephson junction,\cite{Kidwingira:2006aa} and polar Kerr angle\mre{\cite{Xia:2006hd,Komendova:2017cu}} measurements \mre{which} show spontaneous broken time reversal symmetry below \mre{the critical temperature ({\Tc})}.
Additionally, Knight shift\cite{Ishida:1998tx} and SQUID junction\cite{Nelson:2004ga} measurements of the susceptibility \mre{indicate} triplet pairing.
At the same time, seemingly contradictory \mre{or inconclusive} experimental results have left important open questions concerning the detailed structure and coupling of the orbital and spin parts of the order parameter.\mre{\cite{Mackenzie:2017wc}}
\mre{Although studies of {\Tc} under strain do show a substantial increase of the critical temperature, the expected cusp at zero strain is not observed.\cite{Hicks:2014jaa}}
Low energy excitations indicate the existence of vertical line nodes in the superconducting gap,
inconsistent with a $p$-wave order parameter.\cite{Hassinger:2017aa}
Also, the first order nature of the upper critical field ({\Hcii}) at low temperature is suggestive of Pauli limiting,\cite{Yonezawa:2013je}
\mre{and has been interpreted as evidence against equal spin pairing required for $p$-wave superconductivity.\cite{Machida:2008cd}
Alternatively, it was suggested that the simple classification of either spin-singlet or spin-triplet pairing is not appropriate, due to strong spin-orbit coupling in {\SRO}.\cite{Veenstra:2014ka,Zhang:2016bz}
Furthermore, recent work has attributed the suppression of {\Hcii} to so-called interorbital effects rather than Pauli limiting.\cite{Ramires:2016gt}}

Superconducting vortices, introduced by an applied magnetic field, may serve as a sensitive probe of the superconducting state in the host material.
Small-angle neutron scattering (SANS) studies of the vortex lattice (VL) have proved to be \mre{a valuable} technique,
often providing unique information about the superconducting order parameter including gap nodes and their dispersion,\cite{Riseman:1998cz,Kealey:2000ku,Riseman:2000vg,Huxley:2000wh,White:2009jp,KawanoFurukawa:2011hw,Gannon:2015ct}
multiband superconductivity,\cite{Cubitt:2003ip,Kuhn:2016ej}
Pauli paramagnetic effects,\cite{DeBeerSchmitt:2007jt,Bianchi:2008bq,White:2010hu,Das:2012fb}
and a direct measure of the intrinsic superconducting anisotropy ({\Gac}).\cite{Christen:1985wo,Gammel:1994vn,Kealey:2001fi,Pal:2006gi,Das:2012fb,KawanoFurukawa:2013cs,Rastovski:2013hh,Kuhn:2016ej}
The latter quantity may be directly measured by the field-angle-dependent distortion of the VL structure from a regular triangular symmetry.
In London theory,  {\Gac} represents the anisotropy of the penetration depth.\cite{Thiemann:1989uw,Daemen:1992tx}
In Ginzburg-Landau theory it also represents the anisotropy of the coherence length, which can arise from both superconducting-gap and Fermi-velocity anisotropy.
\mre{A determination of %Determining
{\Gac} is particularly relevant} %important
in materials where the upper critical field is Pauli limited along one or more crystalline directions,
since the {\Hcii} anisotropy may differ from the intrinsic superconducting anisotropy.

Here we report SANS studies of the VL in {\SRO} with magnetic fields close to the basal plane in order to investigate the superconducting anisotropy as well as possible effects of Pauli paramagnetism.
In earlier work we found $\Gac \approx 60$ at intermediate fields and low temperature (50~mK).
This significantly exceeds the low-temperature upper critical field anisotropy $\GHcii = \Hcii^{\perp c}/\Hcii^{\parallel c} \approx 20$.\cite{Deguchi:2002bv}
The Fermi surface in {\SRO} consists of three largely two-dimensional sheets with Fermi velocity anisotropies ranging from 57 to 174,\cite{Bergemann:2003dy}
and %in the absence of Pauli limiting
one would expect an upper critical field ({\Hcii}) anisotropy within this range.\cite{Campbell:1988vf,CHANDRASEKHAR:1993wh}
The present work substantially extends the field range of our previous report, and also includes temperature dependent measurements.
The temperature dependent intensity is consistent with gap nodes or deep minima in the order parameter.
No temperature dependence of {\Gac} is observed, but a field driven increase above 1~T indicates multiband superconductivity.
While the discrepancy between {\Gac} and {\GHcii} indicates Pauli limiting,
no direct evidence for Pauli paramagnetic effects on the unpaired quasiparticles in the vortex cores \mre{was} observed.

This paper is organized as follows:
in Section~II we describe the \mre{SANS} experimental details.
Results are presented in Section~III, focusing on the VL configuration and anisotropy, rocking curves, and a determination of the VL form factor.
The implications of our results are discussed in Section~IV, with an emphasis on the superconducting gap structure, Pauli limiting and Pauli paramagnetic effects\mre{,}
and evidence for multiband superconductivity.
A conclusion is presented in Section~V.

%% EXPERIMENTAL DETAILS
\section{Experimental Details}
\begin{figure}
  \includegraphics[scale=1.0]{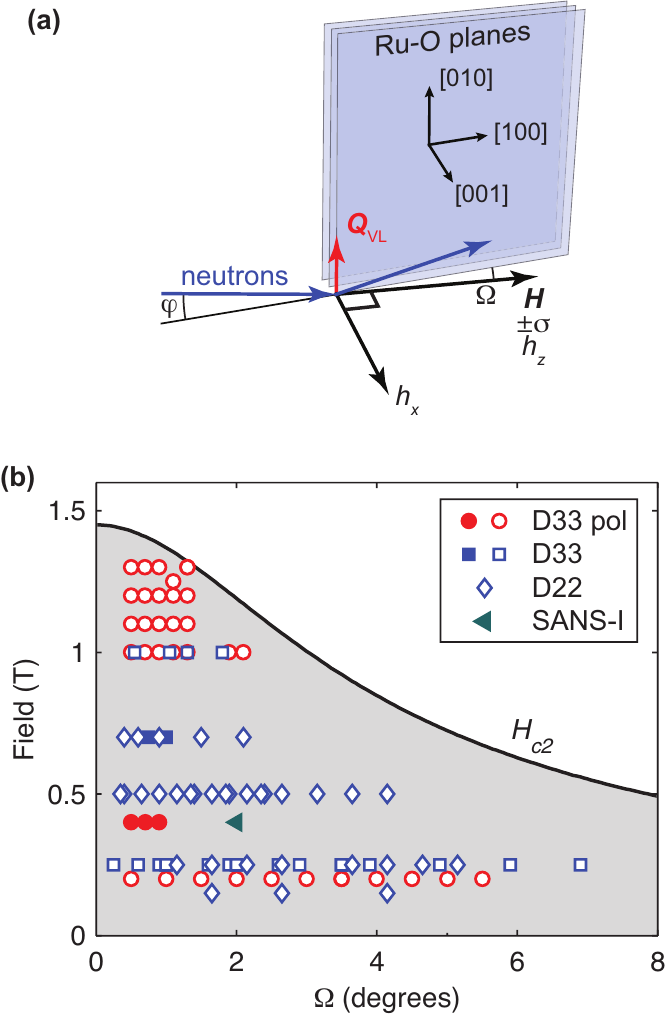}
  \caption{\label{GeomDir}
               Experimental geometry and data overview.
               (a) The coordinate system is defined with $z$ along $\bm{H}$ and $y$ vertical in the Ru-O basal plane.
               \mre{The crystalline axes show the sample orientation used for the ILL experiments.}
               The applied magnetic field $\bm{H}$ is rotated away from the basal plane by an angle $\Omega$.
               Neutron spins ($\sigma$) are parallel or antiparallel to the magnetic field.
               The incident neutron beam is in the $yz$-plane, at an angle ($\varphi$) relative to the field direction.
               The observed VL scattering vector is denoted $\QVLvec$, and $h_z$ and $h_x$ are, respectively, the longitudinal and transverse Fourier component of the field modulation.
               (b) $H$-$\Omega$ phase diagram showing data obtained using the ILL D33, ILL D22, and PSI SANS-I instruments.
               All measurements were performed with $\bm{H} \perp \langle 100 \rangle$, except for the SANS-I data where $\bm{H} \perp \langle 110 \rangle$.
               Solid symbols indicate where measurements were taken at several temperatures; all others were performed at base temperature ($\sim$50~mK).
               The circles (red) indicate measurements using a polarized/analyzed SANS configuration. The upper critical field ($H_{c2}$) is the parametrization from Ref.~\onlinecite{Machida:2008cd}.}
\end{figure}

The superconducting anisotropy {\Gac}  was determined by small-angle neutron scattering (SANS) studies of the vortex lattice (VL). 
These studies simultaneously measure the VL form factor, \mre{$\bm{h}(\bm{Q})$}.
Results from five separate experiments are included in this report, performed at Institut Laue-Langevin (ILL) instruments D22 and D33,\cite{5-42-341,5-42-389}
and Paul Scherrer Institut (PSI) instrument SANS-I. 
The same {\SRO} single crystal with $\Tc = 1.47$~K was used for all the SANS experiments, and was also used in previous work.\cite{Rastovski:2013hh}
The sample was mounted in a dilution refrigerator insert and placed in a horizontal-field cryomagnet.
A motorized $\Omega$ stage rotated the dilution refrigerator around the vertical axis within the magnet, allowing measurements as the magnetic field was rotated away from the basal plane of the tetragonal crystal structure.
The experimental configuration is shown schematically in Fig. \ref{GeomDir}(a).

\begin{figure*}
  \includegraphics[scale=1.0]{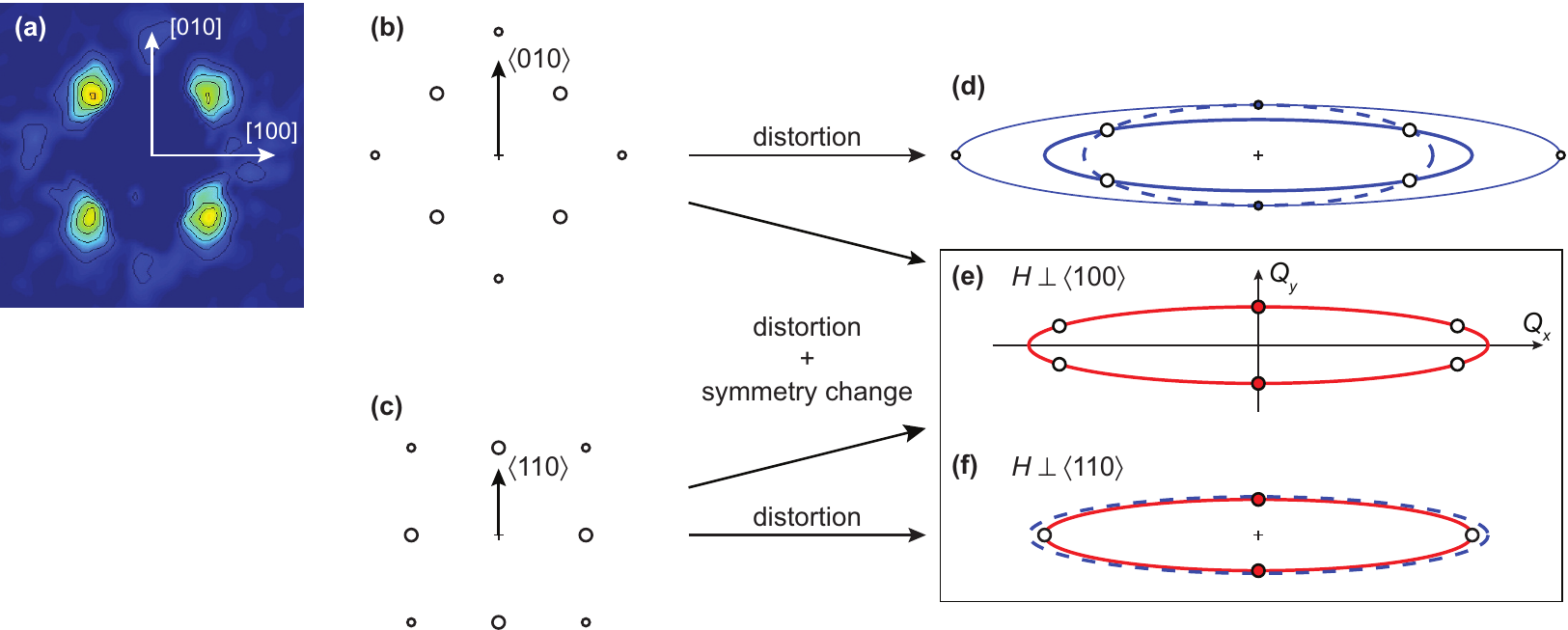}
  \caption{\label{VLconf}
               Possible VL configurations.
               \mre{(a)} SANS diffraction pattern for a field of $25$~mT applied along the [001] axis. A square VL is observed, with Bragg peaks oriented along the $\langle 110 \rangle$ axes.
               The position of the first order (larger circles) and second order (smaller circles) Bragg reflections are shown \mre{schematically} for the two different experimental configurations:
               $\langle 100 \rangle$ vertical (b), and $\langle 110 \rangle$ vertical (c).
               \mre{Rotating} the field towards the basal plane may simply \mre{cause the VL to} distort due to the uniaxial anisotropy (d,f),
               or also undergo a transition from a distorted square to a distorted triangular symmetry (e).
               Here a relatively small anisotropy $\GVL = 6$ was used for illustrative purposes.
               Only VL Bragg peaks on the vertical axis are observed in the SANS experiments (solid symbols), leading to an ambiguity in determining the anisotropy as indicated by the ellipses.
               As discussed in the text, the most likely VL configurations for the two different field directions are shown in (e) and (f).}
\end{figure*}

Measurements were performed with applied magnetic fields between 0.15~T and 1.3~T, applied at angles relative to the basal plane in the range  $\Omega = 0.5^{\circ}$ to $6.9^{\circ}$,
and with temperatures between 50~mK and 1.2~K.
Figure \ref{GeomDir}(b) provides a summary of the measurements.
For the SANS experiments performed at ILL \mre{the} crystalline $\langle 100 \rangle$ axes were horizontal and vertical, while at PSI SANS-I the $\langle 110 \rangle$ axes were horizontal/vertical.
\mre{The two configurations are denoted by respectively $\bm{H} \perp \langle 100 \rangle$ and $\bm{H} \perp \langle 110 \rangle$.
Due to the smallness of $\Omega$ the applied field is also near-parallel to $\langle 100 \rangle$ or $\langle 110 \rangle$.}
The VL was prepared \mre{at low temperature} by first ramping to the desired field ($H$) and rotating to the chosen field orientation ($\Omega$),
followed by a damped small-amplitude field modulation with initial amplitude 50~mT.
This method is known to produce a well-ordered VL in {\SRO}, and eliminates the need for a time consuming field-cooling procedure before each SANS measurement.\cite{Rastovski:2013hh}

The measurements used neutron wavelengths $\lambda_n$ between 0.8~nm and 1.7~nm and a bandwidth $\Delta \lambda_n /\lambda_n = 10\%$.
A position sensitive detector, placed 11-18 m from the sample, was used to collect the diffracted neutrons.
In order to satisfy the Bragg condition for the VL, the sample and magnet were tilted about the horizontal axis perpendicular to the field direction [angle $\varphi$ in Fig. \ref{GeomDir}(a)]. 
Some measurements on the ILL D33 instrument were performed with a polarized/analyzed neutron beam,\cite{Dewhurst:2008gn} as indicated in Fig. \ref{GeomDir}(b) and denoted by ``pol'' in figure legends.
This eliminates the need for background subtraction when measuring the VL spin flip scattering.
For the unpolarized measurements, backgrounds obtained in zero field were subtracted from the data to clearly resolve the weak signal from the VL at high fields.

%% RESULTS
\section{Results}

% Intro
In conventional VL SANS experiments the scattering is due solely to the modulation of the longitudinal component of $\bm{B}(\bm{r})$ in the plane perpendicular to the applied field direction,
denoted $h_z$ in Fig.~\ref{GeomDir}(a).
However, in highly anisotropic superconductors such as {\SRO} there is a strong preference for the vortex screening currents to flow within the basal $ab$ plane.
In this case the associated transverse field modulation ($h_x$) becomes dominant for small, but non-zero,
angles between the applied field and the basal plane.\cite{Thiemann:1989uw,Daemen:1992tx,Amano:2014bja}
It is the relatively large $h_x$ that makes the present VL SANS measurements possible,
since the signal due to $h_z$ for in-plane fields is vanishingly small in {\SRO}.\cite{Rastovski:2013hh}

% VL configuration
\subsection{Vortex Lattice Configuration}
Ideally, $\Gac$ is determined from measurements with the applied field parallel to the crystalline basal plane. 
In this configuration, the primary VL Bragg peaks lie on an ellipse in reciprocal space with a major-to-minor axis ratio given by $\Gac$. 
However, as $h_x$, and thus the VL scattering intensity, vanishes when the field is exactly  parallel to the $ab$ plane such measurements are not possible.
Instead, we determine the VL anisotropy ($\GVL$) with the field applied at an angle {$\Omega$} with respect to the basal plane.
Performing measurements at several angles it is possible to obtain $\Gac = \GVL(\Omega = 0)$ by extrapolation.

The VL distortion due to the uniaxial anisotropy is illustrated in Fig.~\ref{VLconf}.
For fields applied parallel to the $c$ axis a square VL is observed at all fields.\cite{Riseman:1998cz,Riseman:2000vg}
Here the VL is oriented with the primary, first-order reflections along the $\langle 110 \rangle$ axis, Fig.~\ref{VLconf}(a).
As the field is rotated towards the basal plane the square VL is distorted and may also undergo a symmetry change.
The schematics in Fig.~\ref{VLconf}(b,c) show the position of VL reflections for $\bm{H} \parallel \bm{c}$ corresponding to the two orientations of the {\SRO} crystal used in the SANS measurements.
In the first case (b), the sample is rotated around a vertical $\langle 100 \rangle$ axis,
and the horizontal field will therefore always be perpendicular to this direction. We denote this by $\bm{H} \perp \langle 100 \rangle$.
Correspondingly, the second case (c) is denoted by $\bm{H} \perp \langle 110 \rangle$.

\begin{figure*}
  \includegraphics[scale=1.0]{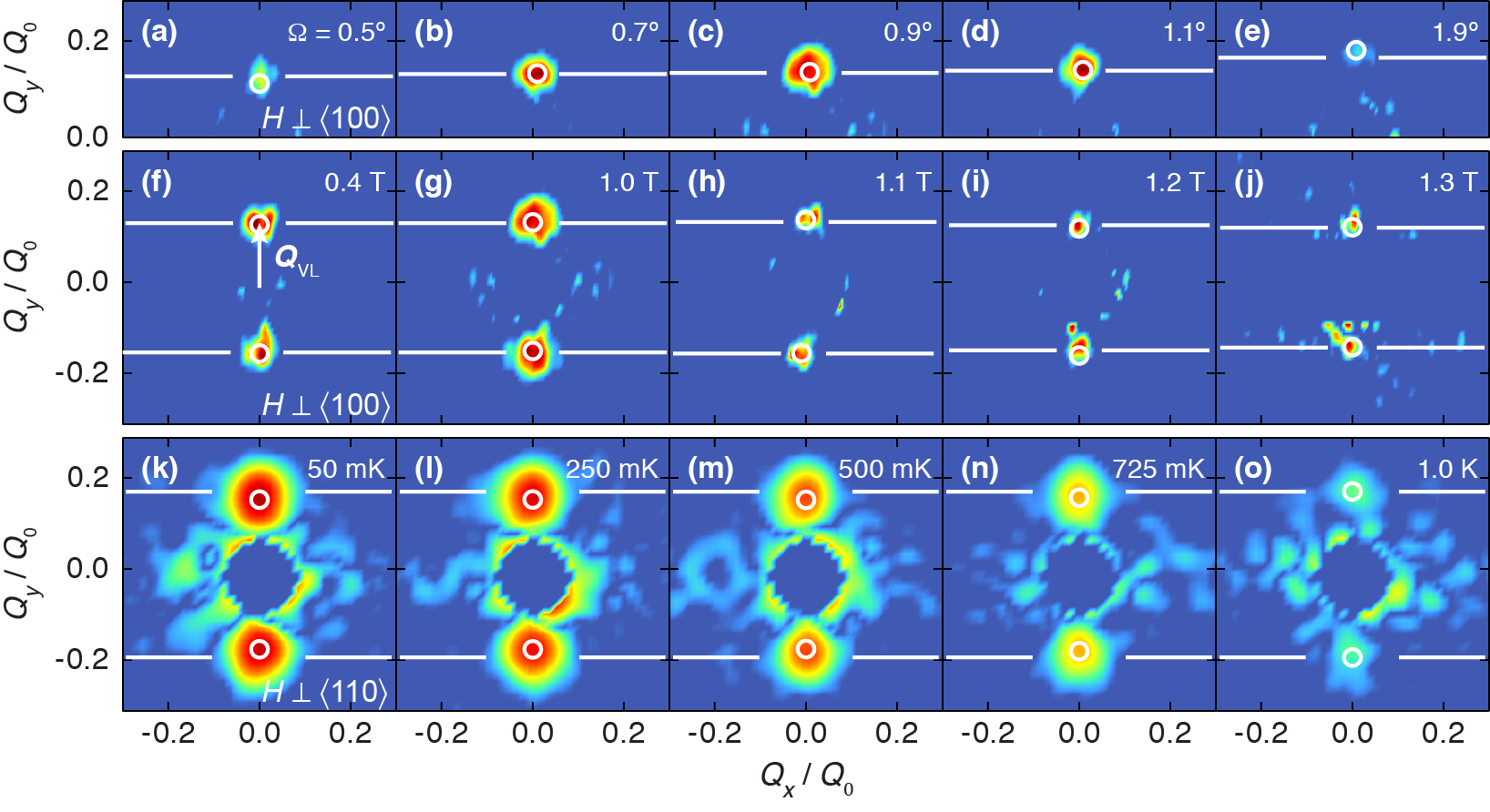}
  \caption{\label{DifPat}
                Vortex lattice diffraction patterns as a function of magnetic field amplitude, rotation angle ($\Omega$) and temperature.
                Positions in reciprocal space (horizontal and vertical axes) are normalized by the scattering vector for an isotropic triangular VL ($Q_0$),
                and data around the direct beam ($Q = 0$) \mre{is} masked off.
                The VL scattering vector is indicated in (f).
                In all cases, the open circles indicate the Bragg peak fitted centers on the detector, while the white lines indicate $Q/Q_0$ values determined from the rocking curves.
                Measurements as a function of rotation angle (a-e, ILL D33) were performed at 1~T ($\bm{H} \perp \langle 100 \rangle$) and 50~mK, for positive $Q_y$ Bragg reflections only.
                The $\bm{H} \perp \langle 100 \rangle$ field sweep (f-j, ILL D33) was performed at $\Omega = 0.9^{\circ}$ and 50~mK.
                Finally, the temperature scan (k-o, PSI SANS-I) was performed at 0.4~T ($\bm{H} \perp \langle 110 \rangle$) and $\Omega = 2^{\circ}$.
                Temperature- and $\Omega$-dependent measurements are shown using fixed intensity scales.
                For the field scan each panel was adjusted separately as the intensity decreases exponentially with increasing $H$.
                Panels (c) and (g) show the same data for $Q_y > 0$.}
\end{figure*}

Fig.~\ref{VLconf}(d-f) show possible VL diffraction patterns that may be obtained as the field is rotated towards the basal plane.
Each vortex carries a single quantum of magnetic flux $\fq = h/2e = 2068$~T$\,$nm$^2$, and as a result the reciprocal space unit cell area is conserved in all cases.
Considering first $\bm{H} \perp \langle 100 \rangle$, the square VL may simply be distorted by an elongation along the $Q_x$ direction and a compression along the $Q_y$ direction (d),
or the distortion may be accompanied by a transition to a distorted triangular symmetry (e).
In the first case, the four first order and the four second order peaks will lie on two separate ellipses, indicated by the solid lines.
In the second case there are six first order peaks lying on the same ellipse.
The same two possibilities exist for the $\bm{H} \perp \langle 110 \rangle$ case, as shown in panels (e) and (f).
In all cases {\GVL} is determined by the major-to-minor axis ratio of the relevant ellipse.

To distinguish between the different VL configurations in Fig.~\ref{VLconf}(d-f) one should in principle measure the position of all the first order Bragg reflections.
However, VL Bragg peaks that are not on the vertical axis (open circles) have scattering vectors almost parallel to $h_x$,
and are effectively unmeasurable as only components of the magnetization perpendicular to the VL scattering vector will give rise to scattering.\cite{Squires:vi}
This introduces an ambiguity as it is not possible to discriminate between (d) and (e) (or between (e) and (f)) based solely on the position of the Bragg peak along the short axis of the ellipse (solid circles).
Experimentally, Bragg peaks are always observed on the vertical axis regardless of the crystal orientation.
This makes the distorted square VL (d) unlikely as the observed peak would correspond to a second order reflection.
For the $\bm{H} \perp \langle 100 \rangle$ case we therefore conclude that the VL undergoes a transition to a distorted triangular VL (e).
For $\bm{H} \perp \langle 110 \rangle$ the magnitude of the scattering vector $\QVLvec$ makes the distorted square configuration (f) the most plausible, as will be discussed in more detail later.

% VL anisotropy
\subsection{Vortex Lattice Anisotropy}
The diffraction patterns in Fig.~\ref{DifPat} show the VL Bragg peaks used to determine the superconducting anisotropy in {\SRO}.
The VL anisotropy is related to the magnitude of the minor axis scattering vector $\QVL = Q_y$ (f).
This is evident from Fig.~\ref{DifPat}(a-e), where $\QVL/Q_0$ increases ({\GVL} decreases) as a constant applied field of $1.0$~T is rotated away from the basal plane.
Here, $Q_0 = 2\pi (2 \mu_0 H/\sqrt{3}\fq)^{1/2}$ is the scattering vector for an isotropic triangular VL,
where we have assumed that the magnetic induction $B$ (vortex density) is equal to the applied magnetic field $\mu_0 H$.
The value of $\QVL/Q_0$ provides a direct measure of the {\GVL} as long as the VL configuration is known.
In the case of a distorted triangular VL (Fig.~\ref{VLconf}(e), $\bm{H} \perp \langle 100 \rangle$) one finds $\GVL = (Q_0/\QVL)^2$.
This relation is slightly modified for the distorted square VL with first order reflections on the vertical axis (Fig.~\ref{VLconf}(f), $\bm{H} \perp \langle 110 \rangle$): $\GVL =  \sqrt{3}/2 \, (Q_0/\QVL)^2$.
The minor difference between the two anisotropies is evident in Fig.~\ref{VLconf}(f) where the ellipse corresponding to the distorted triangular VL is shown by the dashed ellipse.
Finally, if the VL for $\bm{H} \perp \langle 100 \rangle$ was a distorted square and the observed peaks were second order, the anisotropy would be given by $\GVL =  \sqrt{3} \, (Q_0/\QVL)^2$.
In such as case, using the expression for a distorted triangular VL would severely underestimate {\GVL} as shown by the dashed ellipse in Fig.~\ref{VLconf}(d).
However, this would also yield a dramatically different VL anisotropy between the two field orientations,
reinforcing the conclusion that the VL for $\bm{H} \perp \langle 100 \rangle$ does indeed have a distorted triangular symmetry. Finally we note that if one assumes a quantization of $\fq/2$, as reported for mesoscopic rings of {\SRO},\cite{Jang:2011aa} the deduced values for {$\GVL$} would double. We consider this an unrealistic scenario in the present case, with a macroscopic, homogenous sample. It would also cause $\GVL$ to exceed the limit corresponding to a diverging anisotropy, as discussed in more detail in sect.~\ref{MultiDisc}.

In addition to the $\Omega$-dependence discussed above, a field dependence of the VL anisotropy was also found, shown in Fig.~\ref{DifPat}(f-j).
In this case it is necessary to separate the effect of a changing superconducting anisotropy from the increasing vortex density due to the change in the applied field.
To achieve this, the axes in Fig.~\ref{DifPat} have all been normalized by $Q_0$. Plotted in this fashion it is apparent that $\GVL$ increases with increasing field \mre{(peaks moving closer to $Q = 0$)}.
In contrast, no temperature dependence of {\GVL} was observed, as evident from Fig. \ref{DifPat}(k-n), where $\QVL/Q_0$ remains constant within the precision of our measurements.

% Rocking curves
\subsection{Rocking Curves}
\begin{figure}
  \includegraphics[scale=1.0]{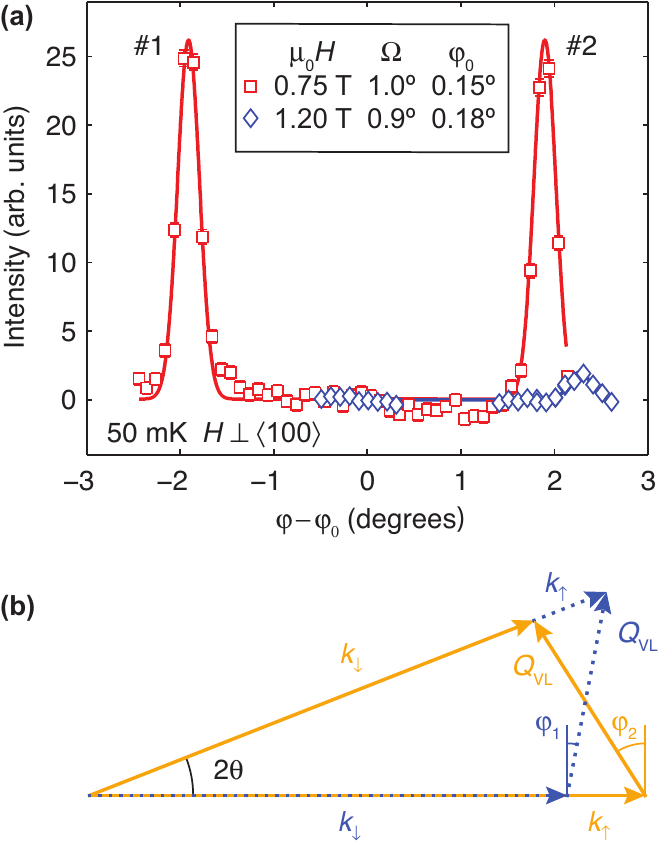}
  \caption{\label{RC}
                (a) Vortex lattice rocking curves at 0.75~T and 1.2~T for the Bragg peaks at positive $Q_y$ \mre{(upper half of detector, see Fig.~\ref{DifPat})} using a non-polarized neutron beam.
                This shows the scattering intensity as a function of sample tilt angle $\varphi$ relative to the rocking curve center, $\varphi_0$.
                Zeeman split peaks due to spin flip scattering are clearly seen,
                while non-spin flip scattering at $\varphi = \varphi_0$ is not observed.
                For 1.2~T only the intensity around $(\varphi - \varphi_0) = 0^{\circ}$ and $2^{\circ}$ was measured.
                \mre{Each peak is} fitted by a Lorentzian.
                (b) Scattering geometry for the two different SF processes:
                Spin-up to spin-down (full lines) and spin-down to spin-up (dotted lines).
                The scattering angle ($2\theta = 2\varphi_0$) is the same in both cases, but different tilt angles ($\varphi_{1/2}$) are required to satisfy the Bragg condition.}
\end{figure}

As an alternative to determining the VL anisotropy from the position of the VL Bragg peaks as discussed above, it is also possible to obtain {\GVL} from the so-called rocking curve.
Figure~\ref{RC}(a) shows the evolution of the scattered intensity as a function of the rocking angle $\varphi$ for a single VL Bragg peak (upper half of the detector) at two different fields of $0.75$~T and $1.2$~T.
In a conventional VL SANS experiment the scattering is due to the longitudinal form factor $h_z$.
As a result, the neutron undergoes non-spin flip (NSF) scattering,
which gives rise to a single maximum in the rocking curve at a tilt angle $\varphi_0 = \QVL/2k_0$ given by Bragg's law in the small-angle limit,
where $k_0 = 2\pi/\lambda_n$ is the nominal neutron wave vector.
However, no NSF scattering is observed for either rocking curve.

Each neutron's spin ($\sigma$) will rotate adiabatically to be either parallel or antiparallel to the magnetic field at the sample position inside the magnet.
The two different directions correspond to different nuclear Zeeman energies and lead to opposite shifts of the neutron wave vector
\begin{equation}
  k_{\uparrow(\downarrow)} = k_0 \, \sqrt{1 \pm \Delta \varepsilon/\varepsilon_0},
\end{equation}
where the subscript in parentheses corresponds to the minus sign in the $\pm \Delta \varepsilon$ term.
Here, $\varepsilon_0 = \hbar^2 k_0^2/2m_n$ and $\Delta \varepsilon = \gamma \mu_N B$, 
where $m_n$ is the neutron mass, $\gamma = 1.913$ is the neutron gyromagnetic ratio and $\mu_N = e\hbar/2m_n = 31.5$~neV/T is the nuclear magneton.
For the fields and neutron wavelengths used in this work $(k_{\uparrow}^2 - k_{\downarrow}^2) \leq 2 \times 10^{-4} \, k_0^2$,
and the difference between $k_{\uparrow}$ and $k_{\downarrow}$ is too small to be observed as a difference in $\varphi_0$ for NSF scattering.
In contrast, spin flip (SF) scattering arising from $h_x \perp \sigma$ leads to two different scattering processes, shown schematically in Fig.~\ref{RC}(b).
Since $\QVL \ll k_{\uparrow/\downarrow}$, the small difference in the neutron wave vectors nonetheless leads to significantly different tilt angles.

The Zeeman splitting of the rocking curve is clearly seen in Fig.~\ref{RC}(a) for 0.75~T.
The 1.2~T data show a qualitatively similar behavior, except that only one of the maxima was rocked through the Bragg condition and the intensity is significantly reduced.
For SF scattering Bragg's law is replaced by
\begin{equation}
  \mp (k_{\downarrow}^2 - k_{\uparrow}^2) - \QVL^2 = 2k_{\uparrow(\downarrow)} \, \QVL \sin \varphi_{1(2)},
\end{equation}
where the subscripts in parentheses correspond to the plus sign.
From the splitting $\Delta \varphi = |\varphi_1 - \varphi_2|$ one thus obtains $\QVL \approx (2 k_0/\Delta \varphi) (\Delta \varepsilon/\varepsilon_0)$.
The results of both methods of determining $\QVL$, and thereby {\GVL}, are indicated in Fig. \ref{DifPat} by the lines (rocking curve) and open circles (detector image).
Within experimental error the two methods agree, and henceforth the average is used.

% Superconducting Anisotropy
\subsection{Superconducting Anisotropy}
\begin{figure*}
  \includegraphics[scale=1]{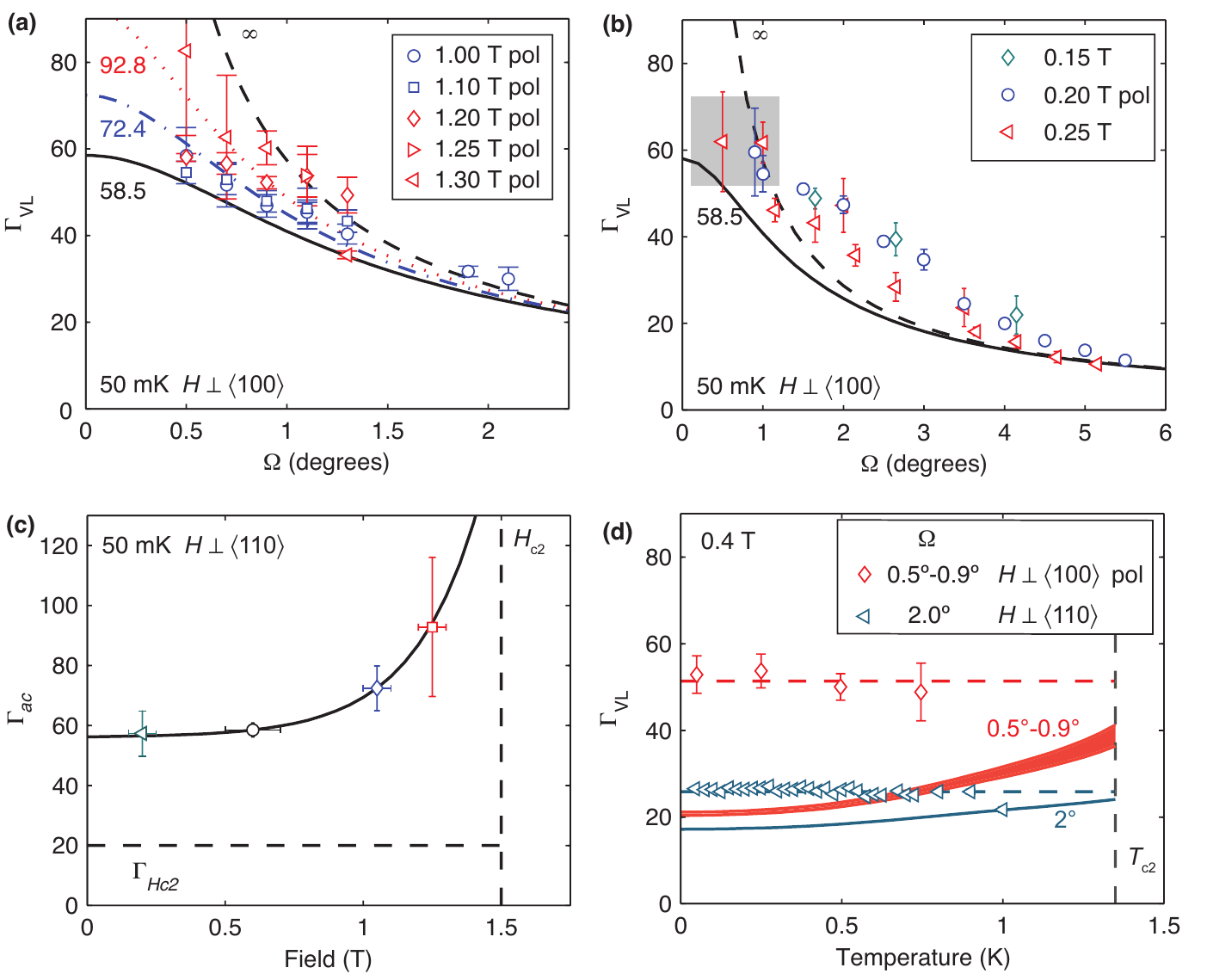}
  \caption{\label{Ani}
                Superconducting anisotropy.
                The vortex lattice anisotropy as a function of field orientation is shown separately for high (a) and low (b) fields.
                Lines are calculated using Eqn.~(\ref{Geq}) and labelled with the corresponding value of \Gac.
                The dashed lines represent a diverging $\Gamma_{ac}$.
                In (a), the lines labelled $\Gac = 72.4$ and $92.8$ are fits to Eqn.~(\ref{Geq}) for the combined 1.0-1.1~T data ($\circ$ and $\square$) and 1.2-1.3~T data ($\diamond$, $\triangleright$, and $\triangleleft$), respectively.
                In (b) the data points in the shaded region were averaged to obtain a \mre{measure of} $\Gamma_{ac}$.
                The solid lines in (a) and (b) show the previously obtained fit to Eqn.~(\ref{Geq}) at intermediate fields of 0.5-0.7~T.\cite{Rastovski:2013hh}
                Panel (c) shows the superconducting anisotropy as a function of field from (a) and (b).
                The horizontal axis error bars indicate the range of fields included in the anisotropy fit.
                The solid line is a guide to the eye.
                Dashed lines indicate the upper critical field and {\Hcii} anisotropy in the zero temperature limit.\cite{Kittaka:2009ub}
                Panel (d) shows the temperature dependence of the VL anisotropy.
                The larger values for $\GVL$ ($\diamond$) represent an average of measurements with $\Omega = 0.5^{\circ}, 0.7^{\circ}$, and $0.9^{\circ}$.
                Anisotropies for data with field orientation $H \perp \langle 110 \rangle$ are calculated assuming a distorted square lattice, as \mre{shown} in Fig.~\ref{VLconf}(f).
                The two horizontal dashed lines are the average of each data set, while the vertical dashed line is  $T_{C2}(\Omega = 0, 0.4 \mbox{~T}) = 1.35$~K.\cite{Deguchi:2002bv}
                The VL anisotropy expected for {$\Gac = \GHcii(T)$} \cite{Kittaka:2009ub} and calculated by Eqn.~(\ref{Geq}) is shown by the line ($\Omega =2^{\circ}$)
                and shaded region ($\Omega = 0.5^{\circ} - 0.9^{\circ}$).}
\end{figure*}
Figure~\ref{Ani}(a,b) shows the VL anisotropy as a function of $\Omega$ for applied fields of $1.0$-$1.3$~T and $0.15$-$0.25$~T, respectively.
These measurements significantly expand previously reported results for $0.5$~T and $0.7$~T.\cite{Rastovski:2013hh}
The low and high field cases are considered separately due to their qualitatively different field dependence.
The high field data are fitted to the equation:
\begin{equation}
  \GVL = \frac{\Gac}{\sqrt{\cos^2 \Omega + ( \Gac \, \sin \Omega )^2}}
  \label{Geq}
\end{equation}
obtained for a 3-dimensional superconductor with uniaxial anisotropy.\cite{Campbell:1988vf}
While {\SRO} is a layered material the coherence length along the $c$ axis, $\xi_c$ = 3.3~nm
is still several times greater than the Ru-O interlayer spacing, and Eq.~(\ref{Geq}) was previously found to provide a \mre{reasonably} good description of the data.\cite{Rastovski:2013hh}
Numerical calculations based on the Eilenberger model %have also validated this simple expression.\cite{Amano:2015kc}
\mre{suggest that this expression slightly underestimates {\GVL}.\cite{Amano:2015kc} However as numerical results are only available for a few values of $\Omega$ we shall rely on Eq.~(\ref{Geq}).}
\mre{As shown in Fig. \ref{Ani}(a), this yields fitted values} of the superconducting anisotropy $\Gac = 72.4 \pm 7.5$ for the combined $1.0$-$1.1$~T data, and \mre{$93 \pm 23$} for the $1.2$ and $1.3$~T data.
Despite the large uncertainty on the fitted values of $\Gac$ the 1.2-1.3~T data ({\GVL}) exceeds the 1.0-1.1~T data over the entire measured $\Omega$ range,
indicating that the superconducting anisotropy increases when approaching $\Hcii$.
This is also consistent with the fitted $\Gac = 58.5\pm 2.3$ obtained previously for fields of 0.5-0.7~T. \cite{Rastovski:2013hh}

\begin{figure*}%
  \includegraphics[scale=1.0]{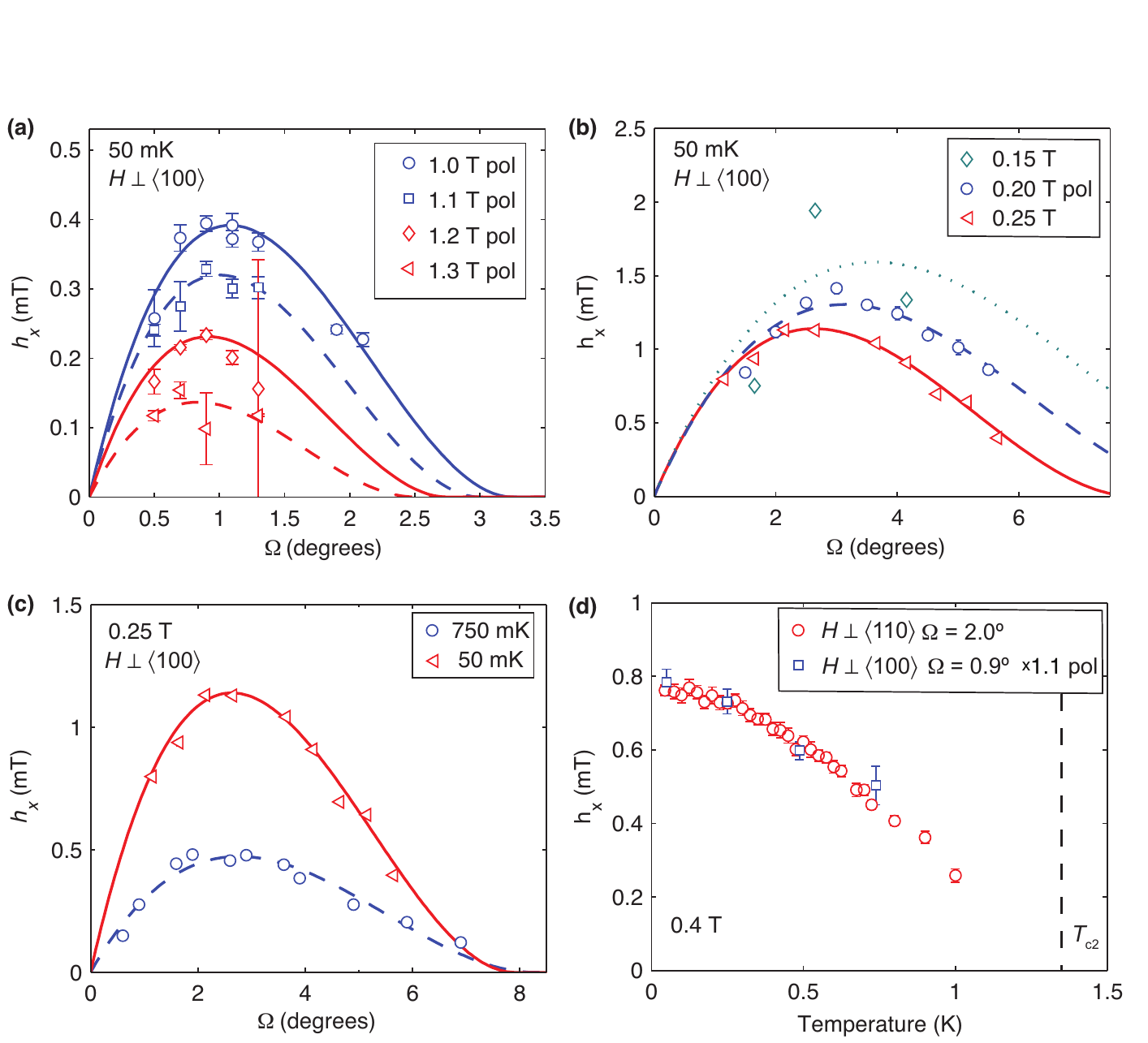}
  \caption{\label{FF}
               Transverse form factor.
               The lines are guides to the eye for each field. Panel (a) shows high field data ($\geq 1$~T.)
               Panel (b) shows data for the low fields most strongly affected by the BLE.
               Both (a) and (b) were measured at base temperature (50~mK).
               Panel (c) compares data taken at $0.25$~T for temperatures of 50 and 750~mK. The 50~mK data is the same as in (b).
               %The 0.25~T, base temperature curve from (b) is shown for comparison (full line).
               The dashed line is obtained by dividing the base temperature curve (full line) by $2.5$.
               Panel (d) shows $h_x$ as a function of temperature.
               Here the $\bm{H} \perp \langle 100 \rangle$ values have been multiplied by $1.1$
               The vertical dashed line indicates $T_{C2}(\Omega = 0, 0.4 \mbox{~T}) = 1.35$~K.\cite{Deguchi:2002bv}}
\end{figure*}

In contrast to the high field data discussed above, the measurements of $\GVL$ in the range of 0.15-0.25~T, shown in Fig. \ref{Ani}(b), \mre{deviate significantly from}
%are not well described by
Eqn.~(\ref{Geq}).
Rather the measured VL anisotropy exceeds the expectation for a diverging $\Gac$.
%, which is colloquially referred to as the ``Buzz Lightyear Effect'' (BLE).
As will be discussed quantitatively later, this Base Line Excess (BLE) discrepancy may be due to multiband superconductivity or a difference between the nominal and actual value of $\Omega$ at low fields.
%For the present we
\mre{In the low field case we instead} obtain a lower limit on $\Gac$ by averaging the measured $\GVL$ for $\Omega \leq 1.2^{\circ}$, indicated by the shaded area in Fig.~\ref{Ani}(b).
The field dependence of the superconducting anisotropy for all magnetic fields is summarized in Fig.~\ref{Ani}(c).
\mre{On close inspection a deviation is also observed at intermediate $\Omega$ at intermediate\cite{Rastovski:2013hh} and high fields, although much less pronounced.
Note that the BLE is {\em not} due an error in the sample alignment,
as measurements of the scattered intensity (Fig.~\ref{FF}) at positive and negative $\Omega$ allow a precise orientation of the crystalline basal plane relative to the field direction.\cite{Rastovski:2013hh}}

The temperature dependence of $\GVL$ at 0.4~T is shown in Fig.~\ref{Ani}(d) for both $\bm{H} \perp \langle 100 \rangle$ and $\bm{H} \perp \langle 110 \rangle$.
The different magnitudes of $\GVL$ are due to differences in $\Omega$. 
For $\bm{H} \perp \langle 100 \rangle$, the data is an average of $\GVL$ for $\Omega$ values of $0.5^{\circ}, 0.7^{\circ}$, and $0.9^{\circ}$,
while \mre{$\bm{H} \perp \langle 110 \rangle$} was measured at a single $\Omega = 2^{\circ}$. 
For the $0.5^{\circ} - 0.9^{\circ}$ data, the sensitivity to changes in $\Gac$ is $\sim 10$.
At larger $\Omega$, the $\GVL$ curves for different  $\Gac$ merge as seen in Fig.~\ref{Ani}(a,b).
At $\Omega = 2^{\circ}$ the sensitivity is therefore reduced, and the uncertainty on $\Gac$ is $\sim 20$.
Within these limits, the data in Fig.~\ref{Ani}(d) indicates that $\Gac$ remains constant as $T$ is increased from base temperature toward $T_{c2}(0.4~$T$) = 1.35$~K. 
In contrast, the $\Hcii$ anisotropy as a function of temperature, $\GHcii(T)$, has been found to increase with increasing temperature from 20 at low temperature to 60 at $\Tc$.\cite{Kittaka:2009ub} 
The expected VL anisotropy if {$\Gac = \GHcii(T)$} was calculated using Eq.~(\ref{Geq}) and shown in Fig.~\ref{Ani}(d). 
In both cases this lies noticeably below the measured {\GVL}.
We note that the VL anisotropies for $\bm{H} \perp \langle 110 \rangle$ were obtained using the expression for a distorted square VL.
From the average of these values we obtain $\Gac = 61_{-12}^{+20}$, in good agreement with the result from the $\Omega$-dependence shown in Fig.~\ref{Ani}(c).
In contrast, values of {\GVL} associated with a distorted triangular VL exceed the limit for a diverging {\Gac}.
As the previously discussed BLE practically vanishes above 0.25~T such a result would be unphysical, supporting the conclusion that for $\bm{H} \perp \langle 110 \rangle$ the VL  has a distorted square symmetry.

% VL Form factor
\subsection{Vortex Lattice Form Factor}
We now return to the measurements of the transverse VL form factor, $h_x$.
The integrated intensity of the Zeeman split Bragg peaks is obtained from rocking curves, such as the one shown in Fig.~\ref{RC}.
Dividing the integrated intensity by the incident neutron flux yields the VL reflectivity, which is related to the form factor by
\begin{equation}
  R = \frac{2\pi \gamma^2 \lambda_n^2 t}{16 \fq^2 \, \QVL} \, |h_x|^2,
\end{equation}
where $t$ is the sample thickness.\cite{Eskildsen:2011jp}
Here, the integrated intensity for the two maxima in the rocking curve are added, as each corresponds to half the incident flux (one direction of the neutron spin). 

For some measurements, a polarized/analyzed SANS configuration was used.
Here, an incident beam polarized parallel or antiparallel to the applied field ($\sigma$ in Fig. \ref{GeomDir}(b)) is scattered by the sample,
and the spin of the outgoing neutrons is selected  using a $^3$He filter before detection.\cite{Dewhurst:2008gn} 
By choosing opposite spin orientations before and after the sample, only neutrons undergoing SF scattering will be measured.\cite{Krycka:2012wa,Wildes:2006ux}
This is an effective method to measure the intensity due to $h_x$, as it suppresses the NSF scattering \mre{background}.\cite{Rastovski:2013hh} 

The form factors obtained in this fashion are shown in Fig.~\ref{FF}.
The same curve shape is used as a guide to the eye for all fields in panels (a-c).
This illustrates how VL SANS measurements are possible within a narrow angular range, with $\bm{H}$ close to, but not perfectly aligned with, the basal plane.
Both the width of the measurement ``window'' and the magnitude of the form factor decreases rapidly with increasing field, as clearly seen for the high field data in Fig.~\ref{FF}(a).  
At low fields, where the BLE is relevant, the curves are found to overlap at low $\Omega$, Fig.~\ref{FF}(b). 
Finally, Fig.~\ref{FF}(c) shows the $\Omega$ dependence of the form factor at $T = 750 \mbox{~mK} = \tfrac{1}{2}\Tc$.
Here the magnitude of the form factor is reduced by a factor of 2.5 relative to the value at base temperature, but otherwise follows the same curve.

The temperature dependence of $h_x$ for two field orientations is shown in Fig.~\ref{FF}(d).
While the form factors are in principle determined on an absolute scale the exact normalization varies slightly from one experiment to another due to differences in sample illumination,
giving rise to minor systematic differences.
In the present case the values for  $\bm{H} \perp \langle 100 \rangle$ were multiplied by $1.1$, to make the form factors overlap at base temperature.
From this\mre{,} one finds that the transverse form factors for the two different field directions follow the same temperature dependence within the precision of the measurements.

Several theoretical models for the form factor exist, with the simplest analytical expressions obtained from the London model.
In this case, the transverse form factor for the observed VL Bragg peaks is given by:\cite{Thiemann:1989uw,Daemen:1992tx}
\begin{subequations}
\begin{eqnarray}
   h_x      & = & \frac{B \lambda^2 \, m_{xz} \, \QVL^2}{d} \label{hx} \\
   d          & = & (1 + \lambda^2 \, m_{xx} \, \QVL^2) (1 + \lambda^2 \, m_{zz} \, \QVL^2) \nonumber \\
               &    & \hspace{4cm} - \lambda^4 \, m_{xz}^2 \, \QVL^4 \label{d}\\
   m_{xx} &  = & \Gac^{-2/3} \sin^2 \Omega + \Gac^{4/3} \cos^2 \Omega \label{mxx} \\
   m_{zz} &  = & \Gac^{-2/3} \cos^2 \Omega + \Gac^{4/3} \sin^2 \Omega \label{mzz} \\
   m_{xz} &  = & (\Gac^{-2/3} - \Gac^{4/3}) \cos \Omega \; \sin \Omega. \label{mxz} 
   \label{FormFactor}
\end{eqnarray}
\end{subequations}
Here $\lambda = (\lambda_{ab}^2 \, \lambda_c)^{1/3}$ is the geometric mean of the penetration depths in the $ab$-plane and along the $c$ axis.
Using the zero temperature literature value $\lambda_{ab} = 152$~nm, \cite{Mackenzie:2003wp} and $\lambda_c = \Gac \lambda_{ab}$ with $\Gac = 58.5$,
we find $\lambda \approx 600$~nm.
This yields $(\lambda \, \QVL)^2 \approx 100 \gg 1$,
and with all $m_{\alpha \beta}$ in Eqs.~(\ref{mxx})-(\ref{mxz}) of at least order unity the form factor expression simplifies to
\begin{equation}
\label{hx2}
  h_x \approx \frac{B}{(\lambda \, \QVL)^2} \frac{m_{xz}}{m_{xx} \, m_{zz} - m_{xz}^2}.
\end{equation}
When using the London model, a correction due to the finite vortex core size is typically included by a Gaussian term \mre{$\exp [-c(\xi \QVL)^2]$},
where $\xi$ is the coherence length and $c$ is a constant of order unity.\cite{Eskildsen:2011iva}
The zero-temperature value for the in-plane coherence length,
estimated from the 75~mT upper critical field parallel to the $c$ axis,
is $\xi_0 = (\fq/2\pi H_{c2}^{\parallel c})^{1/2} = 66$~nm.
For $\mu_0 H = B = 0.4~$T and the measured $\QVL = 0.017$~nm$^{-1}$,
one gets a perfect agreement between the measured and calculated $h_x$ at base temperature with a core cut-off constant $c = 0.81$.
This shows the London model expanded with a core cut-off provides at least a qualitatively accurate estimates of the transverse form factor.
That said, we have previously shown that it does not accurately describe the detailed $\Omega$-dependence of $h_x$.\cite{Rastovski:2013hh}

%% DISCUSSION
\section{Discussion}

% SC Gap structure
\subsection{Superconducting Gap Structure}
The temperature dependence of the VL form factor reflects the structure of the superconducting gap in {\SRO}.
As already discussed, the VL anisotropy remains constant for the measurements with $\bm{H} \perp \langle 110 \rangle$ at 0.4~T and $\Omega = 2.0^{\circ}$, shown in Fig.~\ref{FF}(d).
The only temperature dependence will therefore be through the penetration depth and coherence length.
From Eq.~(\ref{hx2}) one finds that $h_x \propto \lambda^{-2}$, and the form factor is therefore proportional to the superfluid density
\begin{equation}
  \rho_s(\mre{t}) = 1 - \frac{1}{4\pi t} \int_0^{2\pi} \int_0^{\infty} \cosh^{-2} \left( \frac{\sqrt{\varepsilon^2 + \Delta^2(\mre{t},\phi)}}{2t} \right) d\phi \, d \varepsilon,
\end{equation}
where $t = T/\Tc$ is the reduced temperature and the dimensionless superconducting gap $\Delta(\mre{t},\phi)$ is given in units of $\kB \, \Tc$.\cite{Prozorov:2006aa,Eskildsen:2011iva}
The superfluid density decreases with increasing temperature due to thermal excitation of quasiparticles, causing $\lambda$ to increase.
Obtaining information about a nodal gap structure requires measurements at temperatures $T \lesssim \Tc/3$ where the quasiparticle thermal excitation energies are much less than $\kB \Tc$.\cite{Gannon:2015ct}
This is clearly satisfied in the present case, with a base temperature $\sim \Tc/30$.

\begin{figure}
  \includegraphics{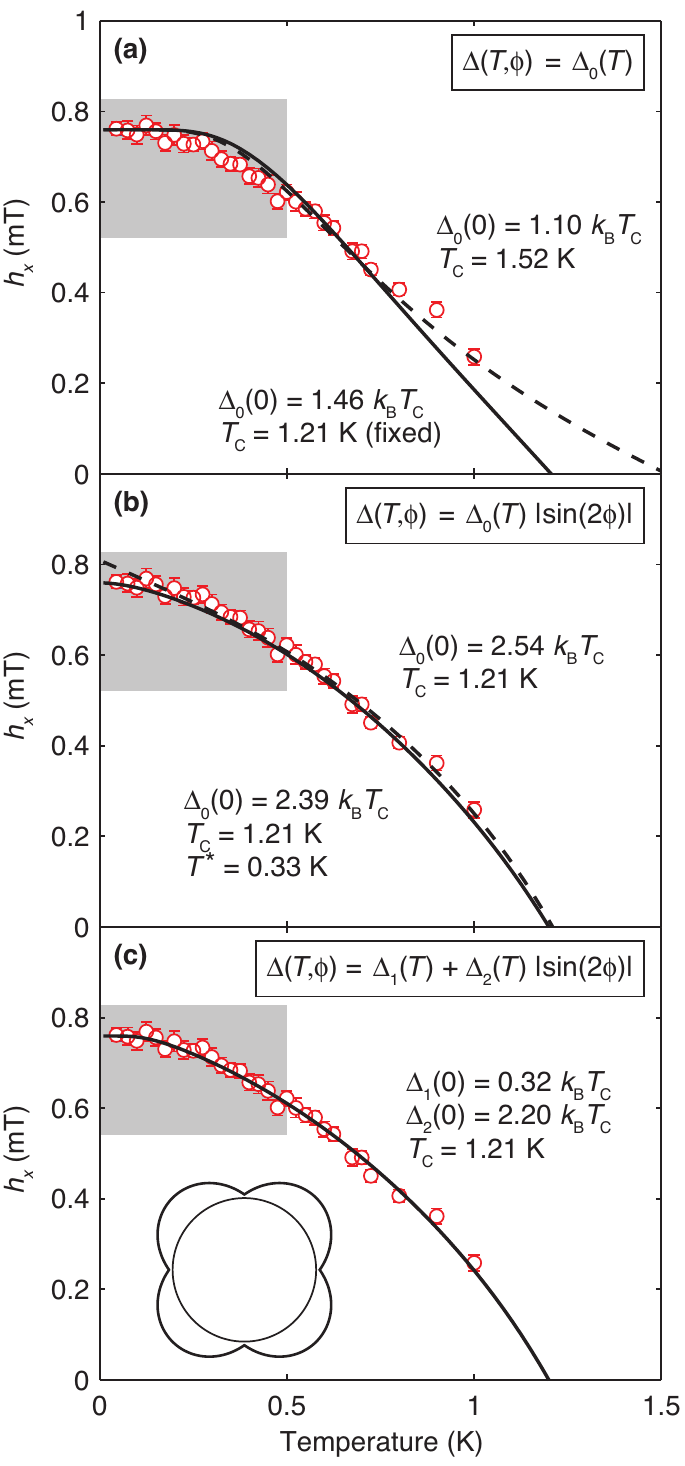}
  \caption{\label{TempDep}
               Fits to from factor temperature dependence.
               Data in all panels is the same as shown in Fig.~\ref{FF}(d) for $\bm{H} \perp \langle 110 \rangle$.
               Fits to \mre{an} isotropic ($s$-wave) gap is shown in (a),
               both with the critical temperature as a free parameter (dashed line) or fixed to $\Tc = 1.21$~K (solid).
               Fits to a nodal gap is shown in (b),
               without (solid line) and with (dashed) non-local corrections.
               The best fit (c) is obtained for a gap with deep, but non-zero minima (schematic).
               All fitting parameters are indicated on the plots.}
\end{figure}

The gap function is separated into temperature-  and momentum-dependent parts $\Delta(\mre{t},\phi) = \Delta_0(\mre{t}) \Delta_k(\phi)$.
For the temperature dependence we use the approximate weak coupling expression,\cite{Gross:1986aa}
\begin{equation}
  \Delta_0(\mre{t}) = \Delta_0(0) \tanh \left( 1.78 \sqrt{\frac{1}{t} - 1} \right)
  \label{GapTDep}
\end{equation}
where $\Delta_0(0)$ is the zero temperature amplitude of the gap.
Replacing $1.78$ by the more accurate $\pi/\Delta_0(0)$ will not affect the conclusion of the following analyis.
In Fig.~\ref{TempDep} we show the results of fits to $h_x$ for different angular dependences of $\Delta_k(\phi)$.
\mre{Here we focus on the difference between the data and the fits} at low temperatures.
In the absence of gap nodes, $\Delta_0(\mre{t})$ in the shaded region is nearly constant,
and  $\rho_s$ will therefore vary little as few quasi particles are excited across the superconducting gap.
In contrast, $\rho_s$ will decrease linearly with temperature near $\mre{t} = 0$ if the gap has line nodes.
We ignore the effect of a temperature dependent coherence length although this would be straightforward to include,
multiplying $\rho_s(\mre{t})$ by the previously discussed core correction and noticing that within the BCS theory $\xi \propto \Delta_0(\mre{t})^{-1}$.
\mre{Including} the core correction would not affect the calculated temperature dependence of $h_x$ in any significant manner at low $T$.

The transverse form factor saturates at low temperatures $< 150$~mK, suggesting a non-vanishing gap on all parts of the Fermi surfaces.
However, a fit to a simple isotropic gap with $\Delta_k(\phi) = 1$ ($s$-wave) provides a poor description of the transverse form factor as shown in Fig.~\ref{TempDep}(a),
regardless of whether the critical temperature is used as a fitting parameter or kept fixed.
Furthermore the fitted $\Delta_0(0)$ is below the lower BCS weak coupling value of $1.76 \kB \Tc$.
A better agreement is obtained for a gap with \mre{line} nodes, Fig.~\ref{TempDep}(b).
Here we have used $\Delta_k(\phi) = | \sin (2\phi)|$ for simplicity.
The differences between this and a $p$-wave order parameter with accidental nodes are expected to be minor.
While the nodal gap in the simplest form varies with temperature all the way to $\mre{t} = 0$ (dashed line),
the London approximation of a vanishing core size relative to the penetration depth will break down in the vicinity of the nodes
and nonlocal corrections should be taken into account.\cite{Kosztin:1997aa,KawanoFurukawa:2011hw}
This leads to a cross-over to a slower temperature dependence below a characteristic temperature $T^* = (\Delta_0(0)/\kB \Tc) (\Tc/\kappa)$,
and yields a transverse form factor
\begin{equation}
  h_x \propto 1 - (1 - \rho_s(\mre{t})) \left( \frac{\Tc + T^*}{\Tc}  \right) \left(\frac{T}{T + T^*}\right). 
\end{equation}
As shown by the solid line in Fig.~\ref{TempDep}(b), this provides a good fit to the data throughout the entire low field region,
although the fitted value of the cross-over temperature ($0.33$~K) is \mre{much} smaller than the theoretical estimate $T^* \approx 1.3$~K when one uses the literature value of $\kappa = 2.3$.
Finally, a comparable but slightly better fit is obtained by an angular dependence of the gap with deep
% (but positive)
minima instead of nodes\mre{:}
$\Delta(\mre{t},\phi) = \Delta_1(\mre{t}) + \Delta_2(\mre{t}) \, | \sin (2\phi)|$, Fig~\ref{TempDep}(c).
The fitted amplitudes for the nodal ($2.39-2.54 \kB \Tc$) or deep minima ($2.52 \kB \Tc$) gaps suggest strong coupling,
\mre{and are in good} agreement with results of scanning tunneling spectroscopy which found $\Delta_0(0) = 350~\mu\mbox{eV} = 2.8 \kB \Tc$.\cite{Firmo:2013dh}
%\mre{From our measurements it is not possible to determine the location of gap nodes or minima.}

The structure of the superconducting gap, and whether this varies between the three Fermi surface sheets,
has been a topic of extensive discussions.\cite{Zhitomirsky:2001aa,Nomura:2005fe,Raghu:2010aa,Wang:2013aa,Scaffidi:2014cx}
For a chiral $p$-wave order parameter gap nodes are not required by symmetry, and in the simplest case the gap is expected to be isotropic.
However, numerous experiments have found evidence for either accidental nodes or deep minima in the superconducting gap from specific heat,\cite{Nishizaki:1999ud,Nishizaki:2000aa,Deguchi:2004aa,Deguchi:2004gp}
penetration depth measurements,\cite{Bonalde:2000wf}
or ultrasound attenuation.\cite{Lupien:2001aa}
Our SANS results are fully consistent with this scenario, although we are not able to determine the location and orientation of the nodes/minima.
More recently, an analysis of specific heat and thermal conductivity measurements put an upper limit on the gap minima $\simeq$~1\% of the gap amplitude.\cite{Hassinger:2017aa}
While the fits in Fig.~\ref{TempDep}(b) and (c) do not allow us to distinguish between actual nodes or deep minima  ($\Delta \chi^2 = 7$\%),
the latter yields a minima-to-amplitude ratio $\Delta_1/(\Delta_1 + \Delta_2) = 0.13$.
This exceeds by an order of magnitude the above mentioned upper limit obtained from measurements of the thermal conductivity.

% PPE
\subsection{\mre{Possible} Pauli Limiting and Pauli Paramagnetic Effects}
The striking difference between {\Gac} and {\GHcii} indicates a strong suppression of the upper critical field in {\SRO} at low temperatures for $\bm{H} \perp \bm{c}$.
This suggests Pauli limiting due to the Zeeman splitting of spin-up and spin-down carrier states by the applied magnetic field.\cite{Clogston:1962aa}
Further support for this comes from the temperature dependence of {\GHcii} which increases towards {\Gac} as $T \rightarrow \Tc$,
indicating that the Pauli limiting of the in-plane upper critical field becomes progressively stronger at lower temperatures. 
In contrast, the lack of a temperature dependence of {\Gac} [Fig.~\ref{Ani}(d)] is consistent with {\Gac} being a measure of the intrinsic superconducting anisotropy arising from the Fermi surfaces.
%Similarly, the previously mentioned magnetic torque measurements also find the coherence length anisotropy to be independent of $T$.\cite{Kittaka:2014kk}

In spin-triplet superconductors the order parameter is most conveniently described in terms of the $d$ vector,
directed along the zero spin projection axis where the configuration of the Cooper pairs is given by $1/\surd 2(\uparrow \downarrow + \downarrow \uparrow$).\cite{Mackenzie:2003wp,Maeno:2012ew,Kallin:2012kx,Kallin:2016bt}
Pauli limiting in the triplet case is therefore only possible if $\bm{H} \parallel \bm{d}$,
which is inconsistent with the chiral superconducting state with $\bm{d} \parallel \bm{c}$ proposed for {\SRO}.\cite{Maeno:2012ew,Kallin:2012kx}
We note, however, that Pauli limiting itself appears to be in disagreement with nuclear magnetic resonance and nuclear quadrupole resonance Knight-shift measurements
(summarized in Ref.~\onlinecite{Maeno:2012ew}), which suggest that the $d$ vector rotates in the presence of a magnetic field such that $\bm{d} \perp \bm{H}$.

Previously, we have used SANS measurements to obtain direct evidence for Pauli paramagnetic effects (PPEs) in superconducting TmNi$_2$B$_2$C and CeCoIn$_5$.\cite{DeBeerSchmitt:2007jt, Bianchi:2008bq, White:2010hu}
In these compounds, a strong coupling to the magnetic field leads to a polarization of the unpaired quasiparticle spins in the vortex cores,
and thus a spatially varying paramagnetic moment commensurate with the VL.\cite{Ichioka:2007kl,Michal:2010aa}
This adds to the orbital field variation in the mixed state,
giving rise to an increase in the total field modulation and hence the longitudinal form factor ($h_z$) with increasing field.\cite{DeBeerSchmitt:2007jt,Bianchi:2008bq,White:2010hu}
Recently, such effects were also found in {\KFA} ($\Gac \simeq 10$) employing a measurement scheme with fields near parallel to the basal plane, analogous with the present work.\cite{Kuhn:2016ej}
In {\KFA} both NSF and SF scattering were observed, and PPEs were inferred from the intensity which deviated significantly from the London model expectation.\cite{Thiemann:1989uw,Daemen:1992tx}

In the present case of {\SRO}, no NSF scattering associated with $h_z$ is observed.
Rather, the transverse form factor decreases monotonically as the applied field is increased, as seen in Fig.~\ref{FF}(a,b) and in our previously published results.~\cite{Rastovski:2013hh}
That said, the London model (including a core correction) does not provide a quantitative description of the data and in particular the narrow range in $\Omega$ where a non-vanishing $h_x$ is observed.\cite{Rastovski:2013hh}
However, numerical solutions to the Elienberger equations that include PPEs do provide a qualitatively accurate description of the measured $\Omega$-dependence of the  transverse VL form factor,
and thereby further support for Pauli limiting in {\SRO}.\cite{Amano:2015kc,Nakai:2015ds}
The numerical work also provides an estimate of the longitudinal form factor, which despite the PPE enhancement remains approximately two orders of magnitude smaller than $h_x$.\cite{Amano:2015kc}
From the rocking curves shown in Fig.~\ref{RC}, we can provide an upper limit on $h_z$.
Using the 1.2~T data as a reference, and estimating the minimum measurable NSF peak size at $\varphi = \varphi_0$ for any practical count time,
we find the longitudinal form factor must exceed $\sim 0.06$~mT to be observed.
The failure to measure the NSF signal from the VL is thus consistent with the numerical calculations.
An estimate of the longitudinal form factor can also be obtained from the experimentally determined magnetization jump at the first-order {\Hcii}, $\Delta M = 0.074 \pm 0.015$~mT.\cite{Kittaka:2014kk}
One expect $h_z \sim \Delta M/6$,\cite{White:2010hu}
indicating that the form factor is just below the value required to be measurable by our SANS measurements.

\mre{In addition to the increase of {\Tc} mentioned in Section~\ref{Intro},
recent measurements of the upper critical field also found that {\GHcii} can be decreased by more than an order of magnitude by strain.\cite{Steppke:2017fb}
This reduction is driven by a dramatic 20-fold increase of {\Hcii} for fields along the $c$ axis, accompanied by a more modest but still noticeable three-fold increase for in-plane fields.
These changes are attributed to a reconfiguration of the Fermi surface and possibly a change in the order parameter.
Complementary SANS studies would be of great interest, in order to explore the strain dependence of {\Gac} and in relation to multiband superconductivity\cite{Ramires:2017fl} as discussed below.}

% Multiband SC
\subsection{Multiband Superconductivity}
\label{MultiDisc}
The superconducting anisotropy determined from our SANS measurements differs dramatically from the upper critical field anisotropy at low temperature, $\GHcii \simeq 20$.\cite{Deguchi:2002bv}
Following our initial report,\cite{Rastovski:2013hh}
this difference was confirmed by magnetic torque measurements performed in fields near parallel to the basal plane, which found a coherence length anisotropy $\xi_a/\xi_c\sim 60$.\cite{Kittaka:2014kk}
We note that while there also is a subtle ($\sim 3$\%) in-plane variation of {\Hcii} at low temperature,\cite{Kittaka:2009ad}
we are not able to determine whether this is reflected in {\Gac}.
This is due to the relatively poor resolution of our SANS measurements where the anisotropies for $H \perp \langle 100 \rangle$ and $H \perp \langle 110 \rangle$ are identical within the experimental error.

A field dependent {\Gac}, such as the one seen in Fig.~\ref{Ani}(c), is characteristic of a multiband superconductor.
The superconducting anisotropy arises from the intrinsic anisotropy of the Fermi surface, $\langle v_{ab} \rangle / \langle v_c \rangle$.
However, for multiband superconductors {\Gac} will be a weighted average of the anisotropy for each band, according to their contribution to the superconducting state.
As the applied field may suppress the superconductivity differently for each band,
it may also change {\Gac}.
A superconducting anisotropy that changes with field is thus a sign of multiband superconductivity,
and has previously been observed in MgB$_2$\cite{Cubitt:2003ip} and \KFA.\cite{Kuhn:2016ej}
We note that while {\Gac} has so far only been found to increase with increasing field, a decrease in the anisotropy would indicate multiband superconductivity by the same argument.

In the case of {\SRO}, the Fermi surface has three bands denoted $\alpha$, $\beta$ and $\gamma$ with anisotropies
$\Gamma_{\alpha} = 116$, $\Gamma_{\beta} = 56.8,$ and $\Gamma_{\gamma} =174$.\cite{Bergemann:2003dy} %\cite{Damascelli:2001tg}
Our results thus suggest that the least anisotropic $\beta$ band is responsible for determining the superconducting anisotropy at low and intermediate fields,
but also that the superconductivity on this band is suppressed above 1~T.
\mre{This agrees with recent inelastic neutron scattering studies which found that the quasi two-dimentional $\gamma$ band, and not the quasi one-dimensional $\alpha$ and $\beta$ bands,
to be primarily responsible for the superconductivity in strontium ruthenate.\cite{Kunkemoller:2017cy}}
\mre{Theoretically, the role of the individual bands and their interplay has been studied extensively}.\cite{Zhitomirsky:2001aa,Nomura:2005fe,Raghu:2010aa,Wang:2013aa,Scaffidi:2014cx,Nakai:2015ds,Huang:2016de}
While there is broad consensus that all contribute to the superconductivity,
different models vary regarding which bands are predicted to be dominant.
However, in most cases the effects of an applied magnetic field have not been considered in detail.
Recently, Nakai and Machida proposed a model for {\SRO} based on a dominant $\beta$ band.\cite{Nakai:2015ds}
While this seems to be in disagreement with a $\Gac > \Gamma_{\beta}$ at high fields,
the model describes the sharp $h_x(\Omega)$ cut-off observed at low fields which is not possible using a single band.\cite{Amano:2015kc}
\mre{A definitive understanding of how the superconductivity in {\SRO} correlates with the individual bands is thus still lacking.}

\begin{figure}
\includegraphics[scale=1.0]{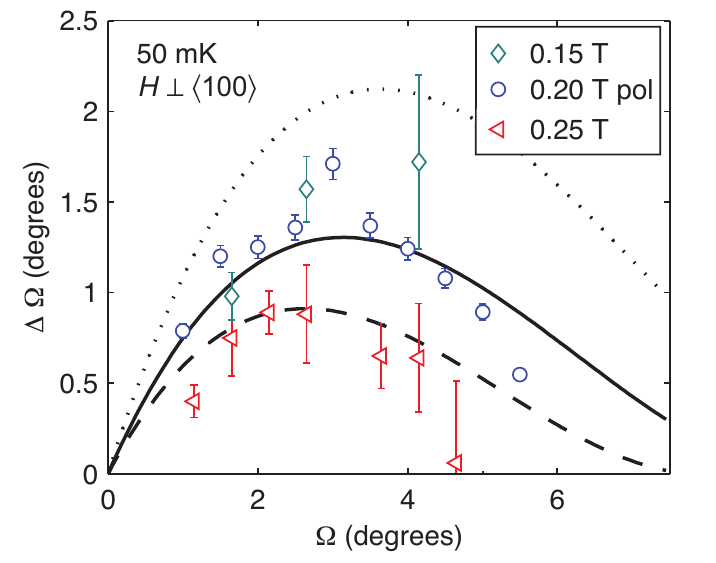}
\caption{\label{DeltaOm}
              Vortex lattice ``misalignment'' as defined in the text.
              %$\Delta \Omega$ vs. $\Omega$. Data points present the separation between the anisotropy data in Fig.~\ref{Ani}(b) and the $\Gac = 58.5$ curve, $\Delta \Omega$.
              The \mre{curves correspond} to those in Fig.~\ref{FF}(b), divided by their corresponding applied fields and multiplied by a common scaling factor.}
\end{figure}

\mre{Finally, we return to the anomalous  $\Omega$-dependence of the VL anisotropy at low fields (BLE), shown in Fig.~\ref{Ani}(b)\mre{,
where {\GVL} clearly exceeds the value expected for a diverging {\Gac} in the range $1.5^{\circ} \leq \Omega \leq 5^{\circ}$.
One possible explanation is provided by the above mentioned model based on a dominant $\beta$ band.\cite{Nakai:2015ds}}
Alternatively, this may be due to a rotation of the vortex direction away from the applied field and towards the basal plane.}
To quantify such an effect we define $\Delta \Omega$ for each data point in Fig.~\ref{Ani}(b),
as the rotation required to shift the measured value of {\GVL} onto the curve corresponding to $\Gac = 58.5$. 
Thus $\Delta\Omega$ would be the ``misalignment'' angle between the nominal and actual VL direction.
This is shown in Fig.~\ref{DeltaOm}.
If a field rotation is responsible for the anomalous values of $\GVL(\Omega)$, $\Delta \Omega$ will be related to the transverse magnetization, $M_x \propto h_x$.\cite{Thiemann:1989uw}
Since $M_x \ll H$, we expect $\Delta \Omega \propto h_x/H$.
The curves in Fig.~\ref{DeltaOm} correspond to those from Fig.~\ref{FF}(b), with each divided by its proper applied field.
After dividing by $H$ all three curves are scaled by the same factor.
\mre{This is in reasonable agreement with $\Delta \Omega(\Omega)$, and we therefore consider a field rotation as the most likely explanation for the BLE.}
Since the ratio $h_x/H$ decreases rapidly with increasing field, the %field
``misalignment'' effect is strongly suppressed at all but the lowest $H$.
\mre{Nonetheless, to fully account for the VL behavior in strongly anisotropic superconductors, a fully three-dimenional treatment is desirable.
Ideally, such a treatment should include realistic material parameters to explain the different VL configurations observed for $\bm{H} \perp \langle 100 \rangle$ and $\bm{H} \perp \langle 110 \rangle$.}
As an aside, we note that demagnetization effects will also provide a negligible change in the vortex lattice direction.
However, from our measurements on {\KFA} we estimate a variation $\leq 0.3^{\circ}$ over the relevant $\Omega$ range.\cite{Pal:2006gi, Kuhn:2016ej}
Additionally, the present experiments were performed using an elliptically cylindrical {\SRO} crystal for which demagnetization effects will be much smaller than for the platelike {\KFA} samples.

\section{Conclusion}
\mre{We have studied the vortex lattice in {\SRO} for fields applied close to the basal plane, nearly parallel to the crystalline $\langle 100 \rangle$ and $\langle 110 \rangle$ directions.
This significantly extends previous SANS measurements which were restricted to low temperature, intermediate fields and a single field rotation axis.
Furthermore, SANS measures the bulk superconducting properties of {\SRO} and allows us to simultaneously address a number of its features.
The use of both spin polarization and analysis in neutron scattering studies of the VL provided an improved signal-to-noise ratio for studies of weak spin flip scattering.}

\mre{Rotating the field towards the basal plane causes a distortion of the square VL observed for  $\bm{H} \parallel \langle 001 \rangle$,
and in the case of $\bm{H} \perp \langle 100 \rangle$ also a symmetry change to a distorted triangular symmetry.
This results in a VL configuration with first-order VL Bragg peaks along the rotation axis for both field orientations.}

\mre{The vortex lattice anisotropy greatly exceeds the upper critical field anisotropy of $\sim 20$ at low temperature, suggesting Pauli limiting.
An increasing anisotropy with increasing field indicates multiband superconductivity, with a value of $\Gac$ between 60 and 100 that suggests a suppression of superconductivity on the $\beta$ band.
In comparison, no temperature dependence of the anisotropy is observed, in striking contrast to $\GHcii$.
We also find that the angular dependence of the VL anisotropy deviates from a simple expression for a uniaxial superconductor, especially at low fields.
A truly three-dimensional model, which includes the salient features relevant to strontium ruthenate, will be required to explain our data over the entire range of fields, field angles and temperatures.}

\mre{Finally, the temperature dependence of the form factor is consistent with either nodes or deep minima in the superconducting gap,
in agreement with recent thermal conductivity measurements.}
\mre{We conclude by noting that a successful model for the superconducting state in {\SRO} must provide an explanation for all the observations summarized above.}

%In conclusion, we have measured the field and temperature dependence of the transverse \mre{vortex lattice} form factor %of the vortex lattice
%in {\SRO} for fields applied close to the basal plane \mre{close to the crystalline $\langle 1 0 0 \rangle$ and $\langle 1 1 0 \rangle$ directions},
%and we have determined the corresponding vortex lattice anisotropy.
%This significantly extends previous SANS measurements restricted to low temperature and intermediate fields\mre{, and represents the first application of both spin polarization and analysis in neutron scattering studies of the VL.
%}.
%The vortex lattice anisotropy greatly exceeds the upper critical field anisotropy of $\sim 20$ at low temperature, suggesting Pauli limiting.
%An increasing anisotropy indicates multiband superconductivity,
%with a value of $\Gac$ between 60 and 100 that suggests a suppression of superconductivity on the $\beta$ band with increasing field.
%In contrast, the anisotropy as a function of temperature is constant.
%Finally, the temperature dependence of the form factor is consistent with either nodes or deep minima in the superconducting gap.

%% ACKNOWLEDGEMENTS
\section{Acknowledgements}
We acknowledge experimental assistance by D. Honecker and J. Saroni as well as useful discussions with M. Ichioka, V. G. Kogan, K. Krycka, and K. Machida.
Funding was provided by the U.S. Department of Energy, Office of Basic Energy Sciences, under Award No. DE-FG02-10ER46783 (neutron scattering),
and by the Japan Society for the Promotion of Science KAKENHI Nos. JP15H05852 and JP15K21717 (crystal growth and characterization).
Part of this work is based on experiments performed at the Swiss spallation neutron source SINQ, Paul Scherrer Institute, Villigen, Switzerland.

%Experiments included:
%2011 ILL D22,
%2012 ILL D33,
%2013 ILL D33,
%2013 PSI SANS-I,
%2015 ILL D33

\bibliography{Sr2RuO4,ILLdoi}

\end{document}